\documentclass{aa}  

\usepackage{graphicx}
\begin{document}

   \title{A new look at microlensing limits on dark matter in the
    Galactic halo}

%   \subtitle{}

   \author{M.R.S. Hawkins
          \inst{1}
          }

   \institute{Institute for Astronomy (IfA), University of Edinburgh,
              Royal Observatory, Blackford Hill,
              Edinburgh EH9 3HJ, UK\\
              \email{mrsh@roe.ac.uk}
             }

   \date{Received September 15, 1996; accepted March 16, 1997}

% \abstract{}{}{}{}{} 
% 5 {} token are mandatory
 
  \abstract
  % context heading (optional)  
   {The motivation for this paper is to review the limits set on the
    MACHO content of the Galactic halo by microlensing experiments
    in the direction of the Large Magellanic Cloud.  This has been
    prompted by recent measurements of the Galactic rotation curve,
    which suggest that the limits have been biassed by the assumption
    of an over-massive halo.}
  % aims heading (mandatory)
   {The paper first discusses the security of the detection efficiency
    calculations which are central to deriving the MACHO content of the
    Galactic halo.  It then sets out to compare the rotation curves
    from various halo models with recent observations, with a view to
    establishing what limits can be put on an all-MACHO halo.}
  % methods heading (mandatory)
   {The main thrust of the paper is to investigate whether lighter halo
    models which are consistent with microlensing by an all-MACHO halo
    are also consistent with recent measures of the Galactic rotation
    curve.  In this case the population of bodies discovered by the
    MACHO collaboration would make up the entire dark matter content of
    the Galactic halo.}
  % results heading (mandatory)
   {The main result of this paper is that it is easy to find low mass
    halo models consistent with the observed Galactic rotation curve,
    which also imply an optical depth to microlensing similar to that
    found by the MACHO collaboration.  This means that all-MACHO
    halos cannot be ruled out on the basis of their observations.}
  % conclusions heading (optional), leave it empty if necessary 
   {Limits placed on the MACHO content of the Galactic halo from
    microlensing surveys in the Magellanic Clouds are inconsistent
    and model dependent, and do not provide a secure basis for
    rejecting an all-MACHO halo.}

   \keywords{dark matter -- gravitational lensing: micro -- Galaxy: halo}

   \maketitle
%
%________________________________________________________________

\section{Introduction}

Over the last three decades it has generally been accepted that studies
of galactic dynamics and the velocity dispersion of galaxy clusters
imply a large component of dark matter which cannot be accounted for by
observable stellar populations, gas and other baryonic material
\citep{t87}.  This finding is supported by results from Cosmic Microwave
Background (CMB) experiments where dark matter is found to make up some
83\% of the mass density of the Universe \citep{d09}.  When Virginia
Trimble wrote her review in 1987 it was already clear that dark matter must
be in a non-baryonic form, and compact bodies were seen as plausible
candidates alongside various supersymmetric particles.  At that time there
was no reason to favour elementary particles over compact bodies,
as little if any direct evidence had emerged to support the existence of
either category of dark matter.

The situation changed dramatically during the 1990s with the first results
from the MACHO collaboration \citep{a96a}.  This well known experiment was
designed to detect the microlensing of stars in the Magellanic Clouds by
compact bodies in the Galactic halo.  The project was a success, and after
5.7 years of observation some 15 microlensing events were observed
\citep{a00}.  This was far more than expected for microlensing by the
stellar content of the halo and disc, but the MACHO collaboration concluded
that the resulting optical depth to microlensing $\tau$ corresponded to a
contribution to the Galactic dark matter halo from compact bodies of only 
$\sim$20\%.  Other groups, notably the EROS and OGLE collaborations,
undertook similar surveys which appeared to put the limit even lower.
Taken together, these results were widely seen as ruling out any
significant component of dark matter in the form of compact bodies of
around a solar mass, and had a major impact on cosmology.

The limitation on the halo fraction in compact bodies derived from the
microlensing of the Magellanic Cloud stars is based on three distinct
components.  Firstly, the events must be detected by searching through
the millions of light curves for the characteristic variation due to
microlensing.  Secondly, the efficiency of the detection procedure must
be estimated to allow a measure of the true optical depth to microlensing
in the Galactic halo.  Thirdly, the Galactic halo must be modelled so
that the optical depth to microlensing can be calculated on the basis
that the dark matter component is composed entirely of compact bodies.
This can then be compared with the observed optical depth.

The idea behind this paper is to re-examine the limits on compact
bodies in the Galactic halo set by the MACHO collaboration and other
groups in the light of recent improvements in the measurement of the
dynamical and structural parameters of the Galaxy.  Modern observations
imply a falling Galactic rotation curve and a relatively light halo.
This has major consequences for the expected number of microlensing
events, which is reduced to as little as 25\% of the previously accepted
value.  The main conclusion from this is that even though there is still
room for discussion on the predicted value for $\tau$, an all-MACHO
halo can no longer be ruled out with any confidence.  Given the
importance of identifying the nature of dark matter, and the current
lack of success in detecting a convincing elementary particle candidate,
it is surely time to look again at the possibility that dark matter is in
the form of compact bodies.

\section{Microlensing limits from the MACHO project}

The idea of detecting dark halo objects by looking for the microlensing of
Magellanic Cloud stars was first suggested by \cite{p86}.  Paczy\'{n}ski
proposed monitoring a few million stars to search for the characteristic
light curve of a microlensing event which would indicate the presence of a
dark halo object.  The proposal was taken up by the MACHO collaboration
which set up a nightly CCD monitoring programme of the Magellanic Clouds
to measure the light curves of around 12 million stars \citep{g91}.  Given
the crowded nature of the star fields, the implementation of a satisfactory
automated data reduction pipeline was in itself a major challenge.
Nonetheless, the first light curve fitting their criteria for a
microlensing event was soon detected \citep{a93}, and the project was shown
to be observationally feasible.  The monitoring programme was completed in
1999 by which time between 13 and 17 events had been detected, depending on
the exact nature of the selection criteria.

\subsection{Detection efficiency}
\label{2.1}

\begin{table*}
\caption{Galactic models for LMC microlensing.}
\label{tab1}
\begin{center}
\begin{tabular}{l c c c c c c c c c c c c}
\hline\hline
 & & & & & & & & & & & & \\
 Model & S & B & F & E & H1 & H2 & H3 & H4 & H5 & H6 & H7 & H8 \\
 & & & & & & & & & & & & \\
\hline
 & & & & & & & & & & & & \\
 $\beta$ & - & -0.2 & 0 & 0 & 0 & 0 & 0.2 & 0.5 & 0.8 & 0 & 0 & -0.2 \\
 $q$ & - & 1 & 1 & 1 & 1 & 1 & 1 & 1 & 1 & 1 & 1 & 1 \\
 $R_{c}$ (kpc) & 5 & 5 & 25 & 20 & 5 & 5 & 5 & 5 & 5 & 5 & 5 & 5 \\
 $R_{0}$ (kpc) & 8.5 & 8.5 & 7.9 & 7.0 & 8.5 & 8.5 & 8.5 & 8.5 &
 8.5 & 8.5 & 8.0 & 8.5 \\
 $\Sigma_{0}$ $(M_{\odot}$pc$^{-2})$ & 50 & 50 & 80 & 100 & 50 &
 67 & 67 & 67 & 67 & 67 & 67 & 67 \\
 $R_{d}$ (kpc) & 3.5 & 3.5 & 3.0 & 3.5 & 2.3 & 2.7 & 2.7 & 3.0 &
 3.6 & 2.5 & 2.6 & 2.6 \\
 $\Theta_0$ (km s$^{-1}$) & 192 & 233 & 190 & 167 & 220 & 220 &
 220 & 220 & 220 & 230 & 210 & 220 \\
 & & & & & & & & & & & & \\
\hline
 & & & & & & & & & & & & \\
 $\chi^2 (<200km/sec)$& 20.9 & 95.8 & 5.7 & 17.0 & 8.4 & 8.6 & 15.7 &
 18.9 & 24.8 & 9.9 & 10.0 & 7.9 \\
 $\tau_{LMC}$ $(10^{-7})$ & 4.7 & 8.1 & 1.9 & 0.85 & 1.64 &
 1.58 & 1.23 & 1.40 & 1.31 & 1.38 & 1.59 & 1.40 \\
 $\chi^2 (<60km/sec)$& 111.2 & 538.0 & 26.5 & 113.2 & 33.4 & 30.6 &
 47.9 & 43.4 & 47.6 & 43.2 & 36.5 & 34.6 \\
 & & & & & & & & & & & & \\
\hline
\end{tabular}
\end{center}
{\it Note.} The predicted values of $\tau$ above may be compared
 with the measurement of $ \tau = 1.2^{+0.4}_{-0.3} \times 10^{-7}$ from
 \cite{a00}.
\end{table*}
Although the detection of a sample of candidate microlensing events formed
the bulk of the enormous effort put into the MACHO project, the next step
of determining the detection efficiency is arguably more critical to the
reliability of the final result.  In an ideal situation the source stars
would be well separated from each other with standard point spread
functions, and the light curves would be adequately sampled with evenly
spaced observations.  The expectation would be that every microlensing
event with given parameters of duration and amplitude would be detected.
In such a case, from a knowledge of the total number of sources being
monitored the frequency of detection of microlensing events is a direct
measure of the optical depth to microlensing, or equivalently the surface
density, of compact bodies in a given mass range in the Galactic halo.

In fact, the data obtained from the microlensing survey in the Magellanic
Clouds were very far from this ideal.  Perhaps the most straightforward
problem was the variable sampling of the light curves due to unfavourable
weather, mechanical failure and other interruptions to the observing
programme.  Gaps in the light curves will result in events being missed
which would otherwise be detected in adequately sampled data.  The MACHO
collaboration's solution to this was to use a Monte Carlo technique to
model the actual epochs of observation and estimate the probability of
missing detections \citep{a01a}.  In these simulations they made the
assumption that each light curve was from a single resolved star.

More serious problems arose from the overlapping of images in the dense
star fields of the Magellanic Clouds.  For example, any star-like image may
actually consist of two or more stars, leading to an underestimation of the
probability of microlensing.  On the other hand, any microlensing
amplification will be reduced as a result of dilution by light from the
other images, and may well result in a colour change if the contaminating
stars are of different spectral types.  This problem is well illustrated
by \cite{a01b} where they show a Hubble Space Telescope image of one of
their microlensing candidates which splits the original single CCD source
into four separate stars.  The MACHO team came to the conclusion that to
attempt to incorporate these effects into more complex Monte Carlo
simulations would not be feasible, and addressed the problem of changes
in photometry from overlapping images in the dense star fields of the
Magellanic by injecting synthetic microlensing events into a sample of
observed light curves \citep{a01a}.  This is a complex process in which the
stellar population must be modelled accurately as a function of position
in the Magellanic Cloud galaxies to allow for changes in the point spread
function from frame to frame, resulting in star images being resolved in
some frames and not in others.  Not only will this reduce the chances
of a microlensing event being detected due to spurious variation, but it
will also produce apparent changes in colour.

There are a number of other effects which must be allowed for in
estimating detection efficiency.  Some of these are listed in Table 4 of
\cite{a01a} and include hard to estimate or unknown parameters.
Particularly problematic is how to allow for the accidental inclusion of
unforeseen types of variable stars such as `bumpers', and the exclusion of
real halo microlensing effects for spurious reasons such as the distortion
of the light curve by the presence of a planetary or binary companion.
This point has been made by \cite{c11} in discussing the OGLE results when
they ``stress the potential difficulty within the evaluation of the
detection efficiency to correctly take into account the risk of excluding
bona fide microlensing candidates''.  The question of whether the detected
events really are microlensing events by compact halo objects has been
the subject of extensive debate \citep{b05b,g05,b05a,e07}, although the
outcome of this exchange seems to support, with minor modifications, the
original claims of \cite{a00}.  Rather than simply removing suspect
microlensing candidates, \cite{b05a} employs a likelihood anaysis to
assign microlensing probabilities to the candidates.  This
a posteriori procedure will in general have the effect of reducing the
observed optical depth to microlensing, and raises the question of a
possible bias as real microlensing events which were inadvertently missed
will not be recovered by subsequent analysis.  As \cite{b05a} points out,
this should be allowed for in the selection criteria, but it is clearly
problematic to anticipate every situation in which a microlensing profile
can be masked.  To illustrate the problem of deciding how to deal with
microlensing events which are discovered after the definition of the
survey sample, and which plausibly might have been contained in it,
we can consider the events detected by one of the MACHO, OGLE or EROS
projects, which although also observed by one of the other two, were
not identified as microlensing events.  Although a posteriori reasons
were usually found to explain this, these events were never included
in other samples because they were assumed to be taken into account in
the detection efficiency calculation.

The possibility that candidate microlensing events might be attributed
to `self-lensing' by lenses in the Magellanic Clouds rather than the
Galactic halo was first suggested by \cite{s94}, and has received much
attention.  This is essentially a part of the calculation of detection
efficiency, and has been addressed in some detail by several groups
\citep{a00,g00,m04}.  Their conclusion was that self-lensing in the LMC
would have only a small effect on the observed optical depth to
microlensing, and certainly could not account for the observed signal.
An interesting paper by \cite{e00} proposes that the LMC is surrounded
by a large microlensing cloud sufficient to reproduce the observed
microlensing signal by lensing background source stars.  The main problem
with this idea is that all known stellar populations in the LMC have too
small a velocity dispersion for such a cloud.  The situation in the SMC
is more complicated as we appear to be observing the galaxy end on,
resulting in a much higher optical depth to self-lensing.  For this
reason, most of the discussion in this paper will be focussed on the
LMC, which in any case dominates in any microlensing statistics.

The MACHO collaboration faced a formidable challenge in confronting the
issues raised above, and tackled it with remarkable thoroughness.  However,
given the importance of their result in ruling out compact bodies as dark
matter candidates, the uncertainties in their procedure cannot be ignored.
Given the lack of knowledge of the underlying starfield, the population
of variable stars and the frequency and effect of binary and planetary
systems it is difficult to see how any firm estimate of the detection
efficiency can be made, let alone a hard lower limit.  It only needs to
drop from around 30\% to 15\% for their favoured halo model to be
consistent with the observed optical depth to microlensing.

\subsection{Halo models}
\label{2.2}

\begin{figure}
\centering
\begin{picture} (0,200) (130,0)
\includegraphics[width=0.5\textwidth]{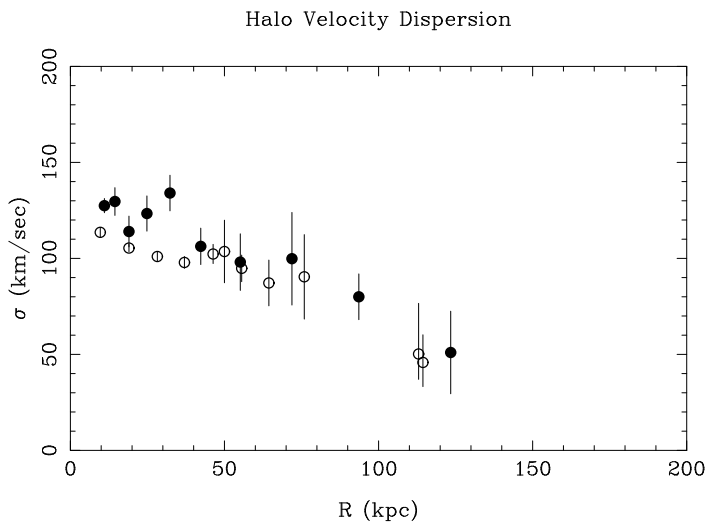}
\end{picture}
\caption{The stellar velocity dispersion as a function of
 Galactocentric distance.  Filled circles are from \cite{b05c} and open
 circles from \cite{d12b}.}
\label{fig1}
\end{figure}

To determine the fraction of the Galactic halo made up of compact bodies,
the sample of microlensing events must be combined with the detection
efficiency to give the observed optical depth to microlensing $\tau$.
This can then be compared with the predicted value of $\tau$ for a chosen
Galactic halo model composed entirely of MACHOs.  It is this choice of
halo model which provides the greatest uncertainty in the limit on dark
matter in the form of compact bodies.  The difficulty which confronted
the MACHO collaboration was that at that time little was known about the
rotation curve, and hence the mass profile, of the Galaxy.  The solution
that they adopted was to define eight model halos incorporating a wide
range of galaxy parameters.  They then calculated the optical depth to
microlensing $\tau$ for each halo on the assumption that it was composed
entirely of compact bodies. By comparing these computed values for
$\tau$ with their observed value they could then determine which, if
any, halo models for the Galaxy could be made up of MACHOs. The model
which they describe \citep{a96a} as a `standard' halo with a core radius
and flat rotation curve, has a density profile of the form:

\begin{equation}
\rho (r) = \rho_0 \frac{R_0^2+R_c^2}{r^2+R_c^2}
\end{equation}

\noindent
where $r$ and $R_0$ are the Galactocentric radius and Galactocentric
distance of the sun respectively, and $R_c$ is the halo core radius.
They also include a set of power law models \citep{e94}, covering a wide
range of halo parameters, including small and maximal discs. The
calculated values of $\tau$ for their models are shown in Table 2 of
\cite{a96a}, and it will be seen that all but two have predicted values
for $\tau$ which are inconsistent with their measured value.  In fact
the authors consider these two Models E and F too extreme to be taken
seriously, although their low optical depths to microlensing are actually
consistent with an all-MACHO halo.  To quote from \cite{a96a} referring
to their model E, ``almost no useful limits $f < 1$ can be placed for
any MACHO mass'', and indeed model E implies a lower value of $\tau$
than that observed by the MACHO collaboration \citep{a96a,a00}.  The
reason the MACHO collaboration rejected Model E was that it ``has an
asymptotic rotation speed of only 83 km s$^{-1}$ and is probably
inconsistent with other estimates of the mass of the Milky Way halo''.
Actually, this figure for the asymptotic rotation speed is close to
recently measured values \citep{s13,b14}.  With regard to their Model F,
\cite{a00} state that it has ``an extremely low mass halo, somewhat
inconsistent with the known Galactic rotation curve''.  It is possible
that at the time of writing this seemed to be the case, but the latest
measurements of stellar velocity dispersion in the outer part of the
Galaxy no longer support this.  Fig.~\ref{fig1} shows recent measurements
of stellar velocity dispersion \citep{b05c,d12b} which imply a falling
rotation curve and a relatively low mass halo.  The actual values of the
rotation speeds are still somewhat uncertain, depending on the values of
the Galactic constants and the velocity anisotropy of the tracer orbits
\citep{b14}, but the declining trend seems clear.  Observations appear
to indicate a radially biassed velocity anisotropy \citep{d12a,r13}
which would favour a lower halo mass.  In fact high values for the Milky
Way halo mass all come from the analyses of the kinematics of the
satellite galaxies including the anomalous object Leo I \citep{b13}.  If
Leo I is taken out of the sample on the grounds that it may not be
gravitationally bound to the Milky Way, or the entire local group is
modelled \citep{p14}, then the kinematics are also consistent with a low
mass halo.  Further support for a low mass halo comes from the analysis
of the Sagittarius stream \citep{g14}, which gives a result free from
many of the assumptions inherent in other approaches.

\begin{figure}
\centering
\begin{picture} (0,200) (130,0)
\includegraphics[width=0.5\textwidth]{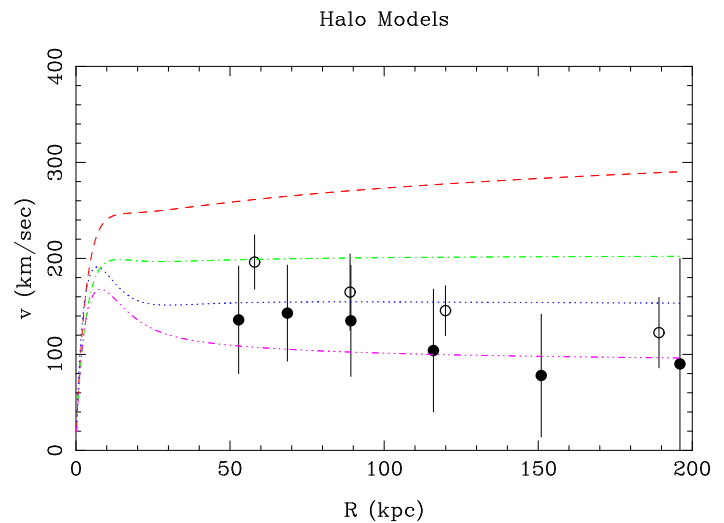}
\end{picture}
\caption{Rotation curves for models for the outer part of the Galaxy.
 The observed rotation curves are shown as filled circles \citep{s13}
 and open circles \citep{b14}.  The red (dashed), green (dot-dash),
 blue (dotted) and purple (dash-dot-dot-dot) curves are for Models
 B, S, F and E respectively from \cite{a96a}.}
\label{fig2}
\end{figure}

For the final analysis of the MACHO microlensing data \citep{a00} just
three halo models were considered, namely S, B and F from Table 2 of
\cite{a96a}.  Model E was not included because, as mentioned above, it
was considered incompatible with observations of the Galactic rotation
curve.  Parameters for these models are given in Table~\ref{tab1},
including model E, together with the optical depth to microlensing
resulting from an all-MACHO halo.  Fig.~\ref{fig2} shows the rotation
curves of these models, and measurements of the circular velocity from
\cite{s13} and \cite{b14} in the outer part of the Galactic halo.
Goodness-of-fit of each model to the data can be assessed from the values
of $\chi^2$ given in Table~\ref{tab1}, with 8 degrees of freedom.  It
will be seen that Models B and S are inconsistent with the data at very
high confidence levels, Model E is marginally rejected, while Model F
provides an adequate fit.  In fact, as quoted above, Model F was rejected
by the MACHO collaboration for being inconsistent with the Galactic
rotation curve as it was known at that time.  The fact that their
preferred Model S appears to be incompatible with recent observations
is undoubtedly a matter for debate, as the calculation of rotation speed
is somewhat model dependent.  However, whether or not Model S should be
seen as a valid model for the Galactic halo, the point to be emphasised
is that all-MACHO Galaxy halos were rejected for reasons which are no
longer valid.

\subsection{Comparison with other surveys}
\label{2.3}

The importance of establishing whether a population of compact bodies
could account for the dark matter in the Galaxy halo prompted other
groups to embark on microlensing surveys in the Magellanic Clouds.  The
EROS collaboration undertook a series of monitoring programmes, the first
of which was based on Schmidt telescope photographic plates.  During the
three seasons of this programme only one ultimately acceptable
microlensing event was detected from the 4 million stars being monitored.
This initial programme was supplemented by a CCD based survey which
monitored four times as many stars in the LMC and resulted in the detection
of a further two microlensing events.  The combined EROS programme was then
analysed by \cite{l00} to put constraints on the MACHO content of the
Galactic halo.  They found that for their `standard' halo model, a solar
mass MACHO component can make up no more than 40\%, at the 95\% confidence
limit. This may be compared with the equivalent figure of 50\% for the
MACHO project \citep{a00}.  The EROS monitoring programme was subsequently
extended in scope as EROS2 and re-analysed by \cite{t07} who restricted
the sample to bright stars with $R < 19.7$, reducing the parent sample of
stars to around 6 million, in the process eliminating all three
previously discovered microlensing events on the basis that they were
observed to brighten again.  Their subsequent analysis failed to identify
any new events, and on this basis they claimed an upper limit of 12\% for
the MACHO content of the Model S halo.

The statistical incompatibility of the MACHO and EROS results has been
the subject of extensive discussion \citep{t07,m10}.  One of the most
obvious differences between the MACHO and EROS experiments is the
respective use of faint and bright star samples.  Only 2 of the 17
MACHO microlensing candidates were bright enough to be included in the
EROS Bright-Stars sample \citep{t07}, which is consistent with the low
detection rate for bright microlensing events in the EROS survey.
Although there are clearly advantages in the restriction of the EROS
analysis to relatively bright stars, it may have resulted in a problem
associated with the resolution of the stellar discs.  A sample of
luminous LMC source stars will contain a high proportion of giants, with
diameters of the order of the Einstein radii of substellar mass lenses.
This can limit the observed microlensing amplitude to as little as a
factor of two \citep{s92}, making the events less likely to be detected
than those associated with smaller, less luminous sources.

As well as including fainter stars in their sample, the MACHO 
collaboration also tended to cover more crowded fields in their survey.
This has raised the possibility that their microlensing detections are
more likely to be caused by self-lensing than in the sparser fields of
the EROS survey.  However, this would contradict results from models of
the LMC which imply that any self-lensing will be small \citep{g00,m04}.
Another potential problem arising from the use of crowded fields is that
poor photometry resulting from blending will result in spurious
microlensing events being selected.  In fact, most of the MACHO
microlensing candidates have withstood subsequent scrutiny and are still
accepted as microlensing events \citep{b05a}.  It is more of an open
question as to how many events have been missed, and whether they have
been correctly allowed for in the detection efficiency calculation.

In reviewing the discrepancy between the MACHO and EROS results,
\cite{m10} concludes that the best way to reconcile them is to drop the
assumption of a homogeneous distribution of microlenses tracing out a
smooth dark matter halo, and to postulate a clumpy halo which the two
surveys sample differently.  This idea has been examined in some detail
by \cite{h06} using N-body simulations of dark matter halos.  They found
that triaxiality and substructure can have major effects on the observed
optical depth to microlensing and event rate.  However, as \cite{m10}
points out, convincing proof of the existence of such structures is still
to be provided.

We have already discussed the difficulties of calculating the detection
efficiency in Section~\ref{2.1}, and this does provide another possible
explanation for at least part of the difference between the MACHO and
EROS detection rates.  A useful way of comparing detection efficiencies
is to look at the statistics of events detected by one group but not by
the other.  There was in fact a considerable overlap between the MACHO
and EROS2 surveys, both in time and fields monitored.  Two of the MACHO
events which fell in this overlap region (MACHO-LMC-18 and 25) were
bright enough to be included in the EROS2 Bright-Stars sample
\citep{t07}, although neither of them was accepted as a microlensing
candidate by the EROS software.  In the case of MACHO-LMC-18 this was
because it was merged with a nearby star \citep{t07}.  These
non-detections will of course have been allowed for in the detection
efficiency calculation, but they also allow a simple Bayesian estimate
of that detection efficiency.  Given that no common detections were
made, no finite figure can be put on this, but it clearly implies a
very low detection efficiency for the EROS2 Bright-Stars sample.
This seems to be at odds with their published detection efficiencies
which are similar to those of the MACHO collaboration.  If the EROS2
collaboration have used detection efficiencies which are too large,
it would go a long way to explaining the discrepancies between the
two groups.

A third group to investigate the MACHO content of the Galactic halo, the
OGLE collaboration, commenced observations in 1996 using broadly the same
techniques as the MACHO and EROS experiments.  The OGLE project went
through a number of phases, the results of which were summarised by
\cite{w11b}.  The survey was divided into `Bright' and `All Stars'
samples to facilitate comparison with the MACHO and EROS results, and
their automated search procedure detected two microlensing candidates
\citep{w11a}, both from the Bright Star sample.  This is broadly in
line with the MACHO collaboration's bright star detections, but the
lack of faint star microlensing candidates is not consistent with the
MACHO result, and hard to explain.  Given that the OGLE All Stars sample
contains some three times as many stars as the Bright sample \citep{w11a},
there should be three times as many microlensing candidates, unless the
fainter events are harder to detect.  This of course is probably the case
due to blending and other magnitude dependent effects, but should be
allowed for in the dection efficiency calculation.  In fact, the
detection efficiencies for the All Stars and Bright samples used for the
analysis were around 15\% and 20\% respectively \citep{w11a}, implying
that magnitude dependent effects were not very important.  This raises
the possibility that the difference in the detection rate between the
MACHO and OGLE programmes could be due to over-estimation of the
detection efficiency for faint stars in the OGLE analysis.

\section{New microlensing limits for the Galactic halo}

\begin{figure}
\centering
\begin{picture} (0,200) (130,0)
\includegraphics[width=0.5\textwidth]{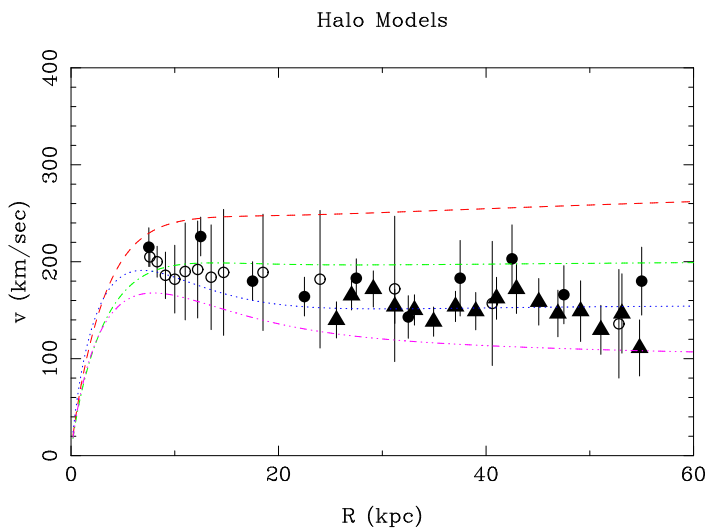}
\end{picture}
\caption{Rotation curves for models of the Galaxy out to 60kpc.
 The observed rotation curves are shown as filled circles \citep{x08},
 open circles \citep{s13} and filled triangles \citep{b14}. The red
 (dashed), green (dot-dash), blue (dotted) and purple (dash-dot-dot-dot)
 curves are for Models B, S, F and E respectively from \cite{a96a}.}
\label{fig3}
\end{figure}

\begin{figure}
\centering
\begin{picture} (0,200) (130,0)
\includegraphics[width=0.5\textwidth]{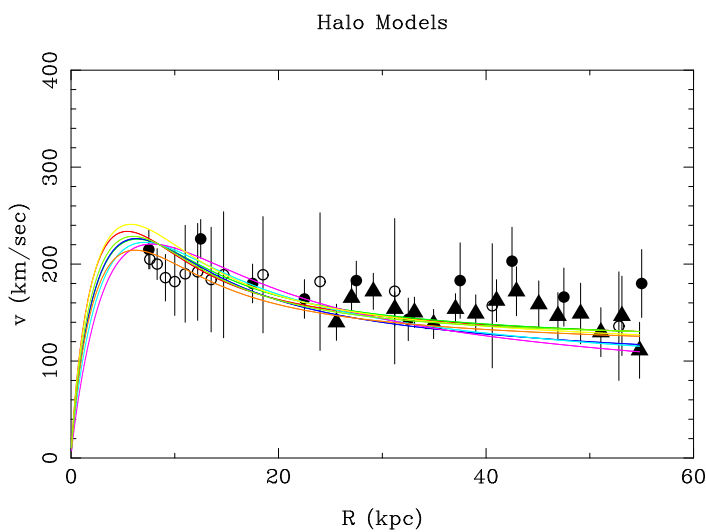}
\end{picture}
\caption{Rotation curves for models of the Galaxy out to 60kpc.
 The observed rotation curves are shown as filled circles \citep{x08},
 open circles \citep{s13} and filled triangles \citep{b14}.  The coloured
 curves are for Models H1 to H8 from Table~\ref{tab1}. }
\label{fig4}
\end{figure}

In Section~\ref{2.2} it has been argued that in the light of recent
observations of the asymptotic rotation speed of the Galaxy, the idea
of an all-MACHO Galactic halo was rejected prematurely.  In this Section
we address the question of whether one can find model all-MACHO halos
consistent with current observational constraints.  For this purpose,
attention needs to be focussed on Galactic structure out to the distance
of the LMC.  Fig.~\ref{fig3} shows recent measurements of the Galactic
rotation curve out to 60kpc together with rotation curves for Models
S, B, F and E from \cite{a00}.  Goodness-of-fit may be assessed from the
$\chi^2$ values given in the bottom line of Table~\ref{tab1}, with 37
degrees of freedom.  It will be seen that Models B, S and E are
inconsistent with the observations at very high significance levels.
Model F provides an adequate fit to the data, with an associated
optical depth to microlensing marginally consistent with an all-MACHO
halo.  However, the Galactic parameters for Model F do not agree well
with recent measurements.  It is nonetheless quite easy to find halo
models with acceptable Galactic parameters that provide a good fit to
the data in Fig.~\ref{fig3}, as well as predicting an optical depth to
microlensing consistent with the MACHO observations and hence an
all-MACHO halo.  Examples of 8 such models are illustrated in
Fig.~\ref{fig4} as coloured lines, with halo parameters given in
Table~\ref{tab1}.  The values of these parameters are within the range
used by \cite{a96a}, and supported by more recent measurements such as
\cite{o01} and \cite{s03}.  As may be seen from Table~\ref{tab1}, the
values of $\tau$ for these halos are consistent with microlensing by a
100\% MACHO halo.  Two of the halos (H4 and H5) are not consistent
with measures of the outer rotation curve, but the remainder provide a
good fit at all Galactocentric distances.  This implies that not
only are all-MACHO halos not ruled out, but that they can actually
provide a better fit to the data than the preferred models of the MACHO
collaboration.

It is important to emphasise that the purpose of this analysis is not to
demonstrate that Model S is inconsistent with the data, but that other
models which imply a low optical depth to microlensing provide as good
a fit.  The question that we address here is whether there is a sound
basis for ruling out an all-MACHO halo from the microlensing observations
of the MACHO collaboration.

\section{Discussion}

The dark matter problem is widely considered to be one of the most
important in physics, and the search for a dark matter particle absorbs
huge resources.  The assumption that dark matter is a particle is
rarely questioned these days, and rests largely on the seminal work
of the MACHO collaboration, and later microlensing surveys in the
Magellanic Clouds.  The most remarkable thing about the results of the
MACHO project is that it discovered a population of compact bodies which
still have no plausible identity.  Whatever the explanation for the
failure of the EROS and OGLE experiments to detect a similar sample of
microlensing events, the MACHO result has withstood scrutiny and needs
explanation.  Additional support for their conclusions comes from the
preliminary results of the SuperMACHO project \citep{r05} which in 2003
alone reported the detection of 10 high quality microlensing events
\citep{b05}.  Although the full analysis of the SuperMACHO results is
yet to be published, the detection rate of microlensing events appears
to favour that found by the MACHO collaboration rather than the very few
events detected by the EROS and OGLE projects.

More supporting evidence for the MACHO result comes from observations
of pixel lensing in M31.  Some 30 candidate microlensing events have
now been reported towards M31 \citep{c10}, but the question of how many
of these are self-lensing events is still a matter of discussion.  The
POINT-AGAPE collaboration conclude that at least 20\% of the halo mass
in the direction of M31 must be in the form of MACHOs \citep{c05}, whereas
the MEGA collaboration \citep{d06} claim an upper limit for a MACHO
halo fraction of 30\%.  They did however concede that their result was
model dependent, and some of the events where hard to explain as
self-lensing due to their position relative to the centre of M31.

The distribution of microlensing candidates across the disc of M31 is an
important way of determining whether the events can be attributed to
self-lensing.  There should be more events from sources in M31 which
are furthest from us, as they would be seen through a greater optical
depth of compact bodies in the M31 halo.  This asymmetry is indeed seen,
but the situation is confused by results that show a similar asymmetry
for variable stars \citep{a04}, which in this case is due to differential
extinction across the disc of M31.  This coincidence has the effect of
blurring the difference between microlensing events and variable stars
which in the absence of extinction would be distinguishable by their
different distributions.  In fact many of the problems associated with
microlensing surveys in the Magellanic Clouds are exacerbated in pixel
lensing projects.  In particular, the detection efficiency calculation
is made much more difficult by the large number of sources in each pixel.

The usefulness of a common halo model to convert optical depth $\tau$ to
MACHO halo fraction has resulted in the widespread adoption of the
MACHO collaboration Model S as a standard for comparison.  In summarising
the OGLE LMC results \cite{w11a} use Model S to calculate the halo mass
fraction, and it is also used by \cite{a96b} together with related models
in the analysis of the EROS results.  In reviewing the results from the
first 20 years of microlensing studies \cite{m10} concludes that a
substantial contribution of compact objects to a standard halo is now
excluded.  However, the `standard' halo to which he explicitly refers is
the MACHO collaboration's Model S. Although a common standard is clearly
useful, it has perhaps obscured the fact that any conclusions rely on
Model S being an accurate description of the Galactic halo.  In fact,
as may be seen Fig.~\ref{fig2} and illustrated in Table~\ref{tab1}, the
asymptotic rotation speed for Model S is not consistent with the observed
rotation curve.  More importantly, in the region out to the distance of
the Magellanic Clouds where any microlensing from bodies in the Galactic
halo will take place Table~\ref{tab1} and Fig.~\ref{fig3} show that
Model S is rejected at high significance level as a fit to the rotation
curve measurements.  By contrast, MACHO Model F is consistent with
measurements of the Galactic rotation curve at all distances, although
its associated predicted microlensing rate for an all-MACHO halo is
$1.8 \sigma$ greater than that found by \cite{a00}.

In addition to the halos from \cite{a96a}, Table~\ref{tab1} also
shows data for 8 halos with parameters in line with recent observations.
All these halos have associated optical depths to microlensing within
around $1 \sigma$ of the value observed by \cite{a00}.  The halos were
selected to be consistent with the recent measurements of the Galactic
rotation curve out to the distance of the LMC, and most of them also
have the low asymptotic rotation speed which is a feature of current
observations.  The values of $\tau$ for all these model halos imply that
they would not be excluded by the results of \cite{a00} as models for
the Galactic halo, and thus that a 100\% MACHO halo cannot be ruled out
for the Galaxy.  The issue here is not how well Model S fits the data,
but whether there are viable halo models which are consistent with a
Galactic halo made up of compact bodies. Given the importance of this
question, it seems premature to abandon this route to the identification
of dark matter, especially in light of the current lack of progress in
detecting a dark matter particle. 

There is still much to be understood about the MACHO content of the
Galactic halo, and in particular the apparent discrepancy between the
results from different groups.  However, the claim that stellar mass
compact bodies have been ruled out as dark matter candidates
on the basis of the Magellanic Cloud microlensing experiments does not
seem to be supportable any more.  A result which has such important
consequences needs to be secure at a high confidence level, and free from
inconsistencies and unquantifiable systematic errors.  The reliability of
the event detection efficiency is fundamental to the confidence that can
be placed on limits to the MACHO content of the Galactic halo.  So far,
no objective way has been found for checking the accuracy of the Monte
Carlo process on which it is largely based.  Moreover, the apparent
inconsistencies between groups in the detection efficiencies for bright
and faint sources that were discussed in Section~\ref{2.3} need to be
clarified.  More important however is the choice of a realistic halo
model that is consistent with observations of the structure and dynamics
of the Galaxy.  We have shown here that models which have been used to
rule out a significant population of compact bodies in the Galactic halo
are not consistent with recent measurements of the Galactic rotation
curve.  On the other hand, halo models that do provide an adequate fit
to the data would, if made up of solar mass compact bodies, produce a
microlensing signal similar to that observed by the MACHO project.
 
\section{Conclusions}

In this paper we re-examine the claim that the Magellanic Cloud
microlensing results of the MACHO collaboration rule out a Galactic halo
composed predominantly of compact bodies.  The MACHO collaboration's
measurement of the optical depth to microlensing, combined with their
preferred Galactic halo model, implies a MACHO fraction of around 20\%.
This result has withstood extensive scrutiny, and the population of
compact objects which it revealed has yet to be satisfactorily
identified.

Recent measurements of the Galactic rotation curve and other structural
and dynamical parameters imply a falling rotation curve and a relatively
light Galactic halo.  We show here that the standard halo model
used by the MACHO collaboration and other microlensing groups is not
consistent with these observations, and consequently cannot be used to
put reliable limits on the MACHO content of the halo.  We then show that
it is easy to find more realistic halo models compatible with the
measurements of Galactic rotation, which if made up of compact bodies
would imply an optical depth to microlensing similar to that found by
the MACHO collaboration.  On this basis we suggest that it is premature
to rule out a Galactic halo composed entirely of compact bodies.  This
result could help to throw light on the current difficulties in detecting
a dark matter particle.

\begin{acknowledgements}
I would like to thank the anonymous referee for some very helpful
suggestions.
\end{acknowledgements}

\end{document}